\def\be{\begin{equation}}
\def\ee{\end{equation}}
\def\bea{\begin{eqnarray}}
\def\eea{\end{eqnarray}}
\def\beg{\begin{align}}
\def\eeg{\end{align}}

\documentclass[a4paper,11pt]{article}
\pdfoutput=1 

\usepackage{jheppub} 

\usepackage[T1]{fontenc} 

\title{\boldmath Quantum and Thermal Fluctuations and Pair-breaking in Planar QED}

\author[a,1]{K. Abhinav,\note{Corresponding author.}}
\author[a]{P. K. Panigrahi,}

\affiliation[a]{Indian Institute of Science Education and Research Kolkata,\\ Mohanpur-741246, India}

\emailAdd{kumarabhinav@iiserkol.ac.in}
\emailAdd{pprasanta@iiserkol.ac.in}

\abstract{Planar quantum electrodynamics, in presence of tree-level Chern-Simons term, is shown to
support bound state excitations, {\it with} a threshold, not present for the pure Chern-Simons 
theory. In the present case, the bound state gets destabilized by vacuum fluctuations. The bound 
state itself finds justification in the duality of the theory with massive topological vector
field. Thermal fluctuations further destabilize this state, leading to smooth dissociation at
high temperatures. Physical systems are suggested for observing such a bound state.}

\begin{document} 
\maketitle
\flushbottom

\section{Introduction}
Planar Chern-Simons (CS) electrodynamics yields an exact, weakly-bound, particle-antiparticle 
bound state (exciton), both in the relativistic \cite{H} and non-relativistic \cite{H2} regimes. Summation
of the bubble diagrams leads to this non-perturbative effect, wherein the binding energy \cite{H},

\be
\epsilon\approx\exp\left(-4\pi\frac{\vert m\vert}{m}\mu/e^2\right),\label{N1}
\ee
with CS coefficient $\mu$, fermion mass $m$ and coupling strength $e$. It is strikingly
similar to the gap of superconductivity
$\vert\Delta\vert\approx2\omega_D\exp\left[\frac{1}{\pi\nu(0)\lambda}\right],~~~\lambda<0$
\cite{Schakel},
where $\omega_D$ is the Debye frequency, and $\nu(0)$ is the density of states. Pure CS QED
is devoid of any dynamics, and only affects the statistics of the interacting particles \cite{H2}.
Coupling to the matter field leads to kinetic terms for the gauge field, which are sub-dominant
in the low-energy limit, being second order in derivative. However, originating from the vacuum-fluctuations
of the matter particles, the same is expected to destabilize the above bound state. Here, we
investigate the role of quantum and thermal fluctuations in pair-breaking through the
non-perturbative Schwinger-Dyson equation (SDE) \cite{SDE,SDE1,IZ}, from the analytical structure of the
full quantum propagator. 
\paragraph*{}In addition to the Maxwell Lagrangian, the presence of the CS term,

\be
\mathcal{L}_{CS}:=\frac{\mu}{2}\epsilon^{\mu\nu\rho}a_{\mu}\partial_{\nu}a_{\rho},\nonumber
\ee
at tree level, makes the theory `massive', while preserving the gauge-invariance of the field
$a_\mu$ \cite{TopM1,TopM11,TopM2,TopM22,Red,Red1}. Consequently, the spins of both gauge ($\mu/\vert\mu\vert$) and
fermion ($m/2\vert m\vert$) emerges \cite{TopM2,TopM22,Boy,Boy1}, which are dependent on the fermion mass
$m$. The topological CS term, that breaks parity, can arise through quantum corrections, owing
to interactions with both bosonic \cite{pkp02} and fermionic \cite{TopM1,TopM11,TopM2,TopM22,Red,Red1,Boy,Boy1,Appel,H}
fields, containing parity-breaking terms. In 2+1 dimensions, the corresponding form factors
logarithmically diverge at momentum equal to $2m$. Consequently, the pole of the full propagator 
from SDE, leads to a physical {\it bound} state just below the two-fermion threshold. The above
1-loop result, though approximate, is justified by large-N arguments \cite{Ren,Inf}, and also 
as the CS contribution does not arise beyond 1-loop \cite{CH}.
\paragraph*{}Such quantum corrections, in general, additionally induces gauge dynamics, through
the kinetic part $\propto F_{\mu\nu}F^{\mu\nu}$. This induces vacuum fluctuations in the gauge
sector, reflected in the pole structure of the full gauge propagator, for the {\it same} value
of the momentum. In this paper, we investigate the formation and subsequent stability of this
bound state, in presence of both vacuum and thermal fluctuations. In the following,
Section I deals with the bound state in presence of gauge dynamics, with the corresponding vacuum
fluctuation shown to impose a finite parametric threshold for the existence of the topological
bound state. This threshold is justified through the duality of the system with massive topological 
vector fields. The effect of thermal influence is dealt with in Section II, depicting smooth
dissociation of the bound state at sufficiently high temperature, through generalization of
existing results. We conclude with discussions and remarks, pointing-out possible physical
realization of such exotic states.

\section{Bound state in presence of vacuum fluctuations}
\begin{figure}
\includegraphics[width=5 in]{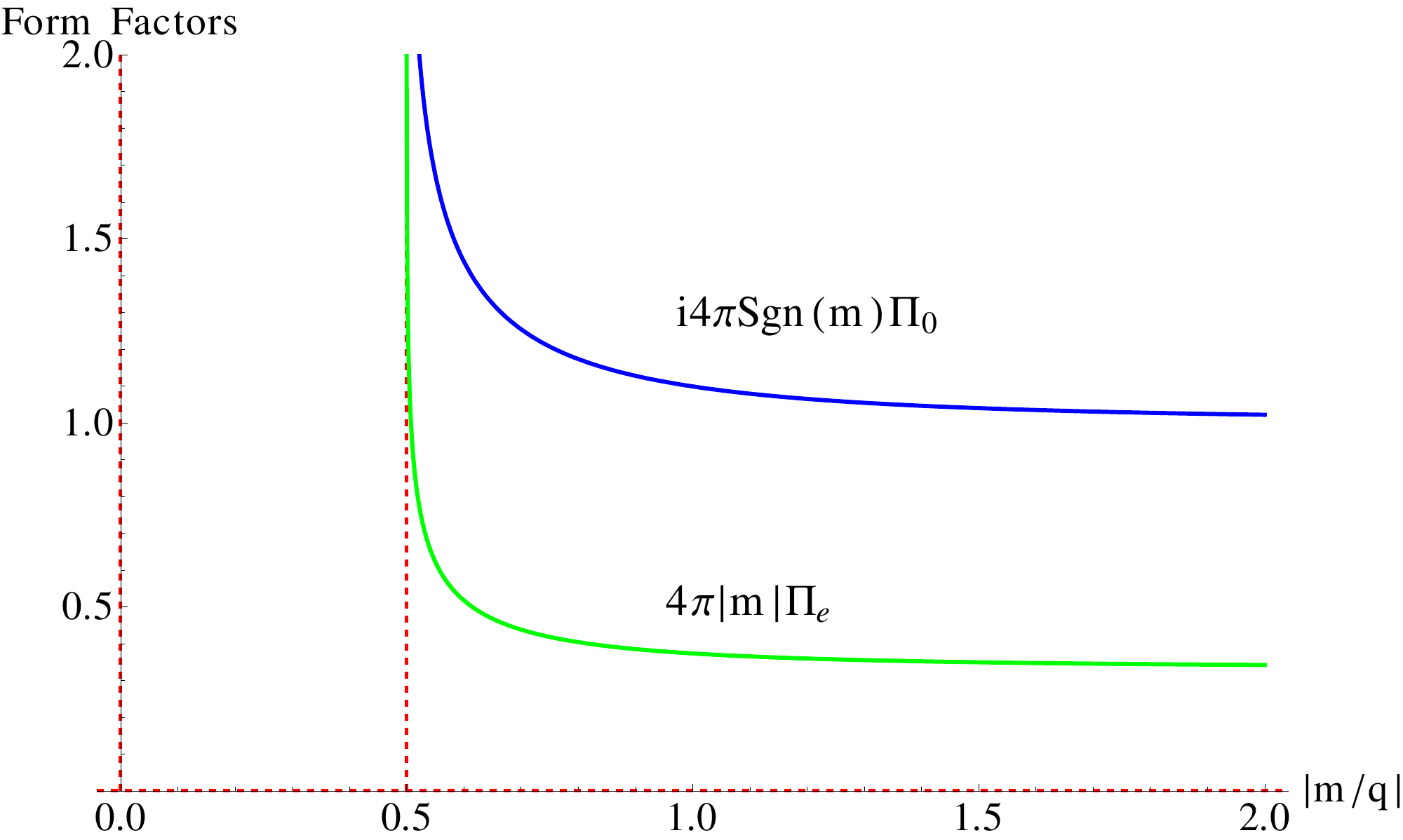}
\caption{Plots of even and odd form factors of vacuum polarization tensor. Both the discontinuities,
shown by red dashed lines, appear at two particle threshold, {\it i.e.}, $\vert q\vert=2\vert m\vert$.
Here $e^2=1$.}\label{f1}
\end{figure}

Planar QED is defined by the Lagrangian,

\bea
\mathcal{L}&=&\bar{\psi}(x)\left(i\gamma^{\mu}\partial_{\mu}-m-\gamma^{\mu}a_{\mu}(x)\right)\psi(x)-\frac{1}{4}F_{\mu\nu}F^{\mu\nu}+\frac{\mu}{2}\epsilon^{\mu\nu\rho}a_{\mu}\partial_{\nu}a_{\rho},\label{1}\\
\gamma^\mu&=&(\sigma_3,i\sigma_1,i\sigma_2),~~~\hbar=c=1,\nonumber
\eea
with coupling constant $e$, where the Dirac algebra is defined by Pauli matrices \cite{TopM1,TopM11,TopM2,TopM22,H}.
One can employ the derivative expansion scheme \cite{DE,DE1,DE2,DE3}, owing to smallness of the coupling constant 
with respect to the corresponding momentum scale. The dominant non-zero contribution comes at the second
order, identified by the vacuum polarization tensor,

\bea
\Pi^{\mu\nu}(q)&=&ie^2T_r\int_p\left[\gamma^{\mu}S(p_+)\gamma^{\nu}S(p_-)+\gamma^{\mu}\frac{\partial}{\partial p_{\nu}}S(p)\right];\nonumber\\
\int_p&=&\int\frac{d^3p}{(2\pi)^3},~~~S(p)=\left[\gamma^{\mu}p_{\mu}-m\right],\nonumber\\
p_{\mu}&=&i\partial_{\mu},~~~p_\pm=p\pm\frac{q}{2},\label{3}
\eea
modulo the normalization owing to the non-interacting fermionic contribution, with the
Dirac trace eliminating the first order term. Here, $q$ is the external momentum of the
gauge field. The second term in the integrand represents the Got\'o-Imamura-Pradhan-Schwinger \cite{GI,Pradhan,Schwinger}
term, utilizing current as the limit including a gauge invariant exponential \cite{H}, which
essentially regularizes the linear UV divergence of planar vacuum polarization \cite{Ren}.
Hence, there is no need of additional gauge-invariant regularization ({\it e.g.}, Pauli-Villars). 
\paragraph*{} The evaluation of the above integral is straight-forward, which can be
carried out both in Minkowski space (using Schwinger parametrization) \cite{G} and
in Euclidean space (using Feynman's trick) \cite{NA3,Rao,Rao01} followed by continuation
back to the Minkowski space, both yielding the result:

\bea
\Pi^{\mu\nu}(q)&\equiv&\Pi_e(q)\left(q^{\mu}q^{\nu}-\eta^{\mu\nu}q^2\right)+\Pi_o(q)\epsilon^{\mu\nu\rho}q_{\rho},\nonumber\\
\Pi_e(q)&=&\frac{e^2}{4\pi}\left[\frac{1}{|q|}\left(\frac{1}{4}+\frac{m^2}{q^2}\right)\log\left(\frac{2\vert m\vert+\vert q\vert}{2\vert m\vert-\vert q\vert}\right)-\frac{|m|}{q^2}\right],\nonumber\\
\Pi_o(q)&=&-i\frac{m}{4\pi}\frac{e^2}{|q|}\log\left(\frac{2\vert m\vert+\vert q\vert}{2\vert m\vert-\vert q\vert}\right).\label{4}
\eea
The parity-odd contribution $\Pi^{\mu\nu}_o(q)$ is special to 2+1 \cite{Red,Red1}, absent
in 3+1, which arises due to non-zero Dirac trace ($Tr_D$) of three gamma
matrices. This is the induced Chern-Simons contribution at loop level, including
the topological L\'{e}vi-Civita tensor. The parity-even contribution $\Pi^{\mu\nu}_e(q)$
is responsible for wave-function renormalization. The plots of both the form factors
$\Pi_{e,o}(q)$ are shown in Fig. \ref{f1}, with the well-known singularities at the
two-particle threshold.

\begin{figure}
\centering 
\includegraphics[width=5 in]{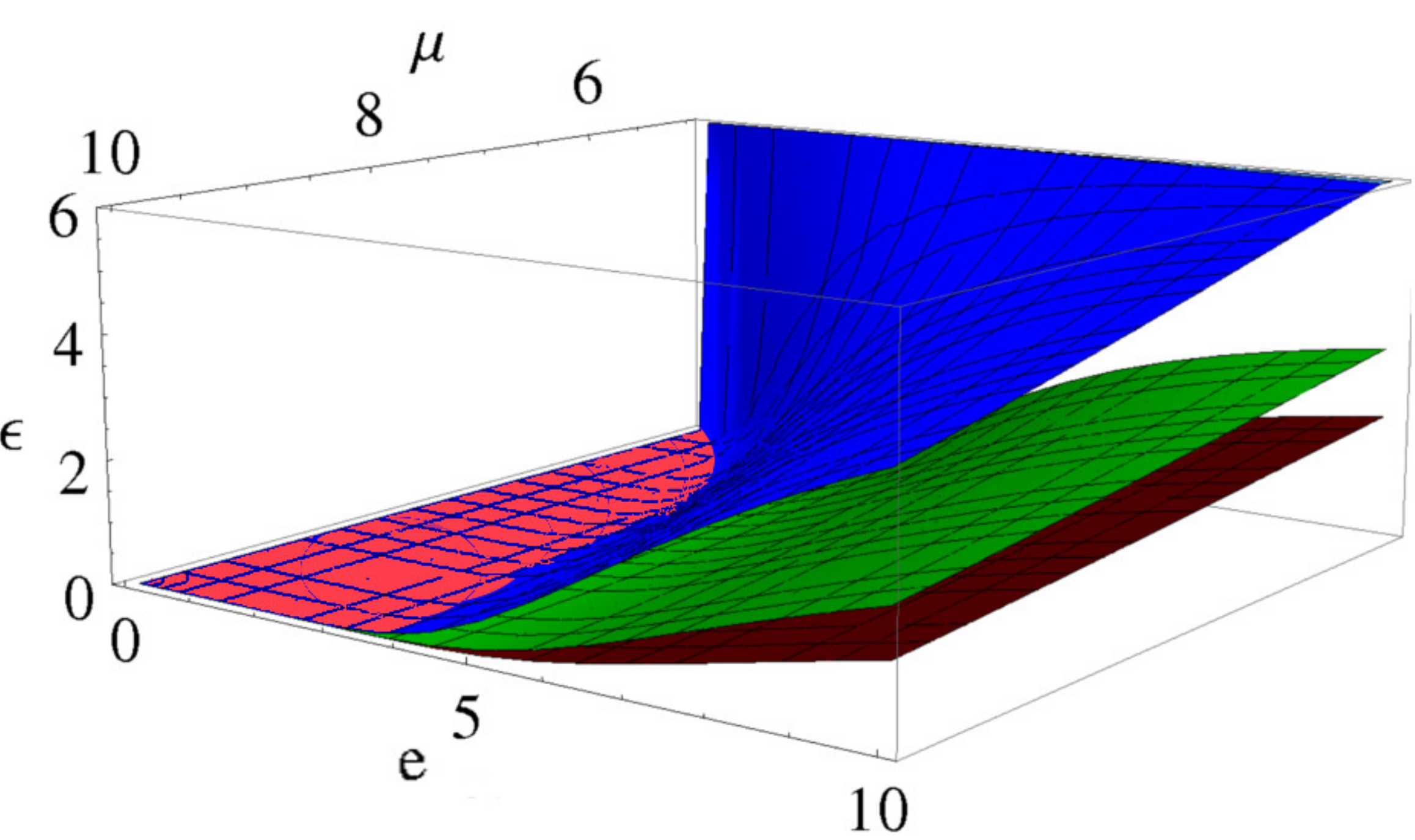}
\caption{Plots of BE $\epsilon$ against interaction strength $e$ and the CS coefficient $\mu$, for $\vert m\vert$ being $1$ (maroon), $1.5$ (green) and $2.5$ (blue) respectively (in natural units). Suitable regions for bound state formation (small $\epsilon$) exist for all three values, marked in red color, signifying the importance of the condition $\mu>2\vert m\vert$.}\label{f2}
\end{figure}

\paragraph*{} It is to be mentioned here that the results in Eqs. \ref{4}, which are valid for 
$q^2<4m^2$, are adopted as we are interested in pole(s) of the gauge propagator just {\it below} the
two-fermion threshold. For $q^2>4m^2$, a branch-cut opens up owing to the singularity at $q^2>4m^2$,
resulting into the replacement:

\begin{equation}
\log\left(\frac{2\vert m\vert+\vert q\vert}{2\vert m\vert-\vert q\vert}\right)\rightarrow\log\left(\frac{2\vert m\vert+\vert q\vert}{2\vert m\vert-\vert q\vert}\right)-i\pi,\nonumber
\end{equation}
in Eqs. \ref{4}.
\paragraph*{}In order to obtain the full gauge propagator, we consider the 1-loop SDE.
The SDE, obtained by setting the variation of the generating functional
$Z:=\int\mathcal{D}[\{\phi\}]\exp[i\int_x\mathcal{L}]$ with respect to the constituent fields
($\{\phi\}$s) equal to zero, is the equation of motion for the particular field
\cite{IZ}. In case of gauge fields, considering the Lagrangian of Eq. \ref{1}, the same turns
out as,

\be
\left[G_F^{\mu\nu}(q)\right]^{-1}=\left[G_F^{(0)~\mu\nu}(q)\right]^{-1}+\Pi^{\mu\nu}(q)-\frac{1}{\xi}q^{\mu}q^{\nu},\label{5}
\ee
up to 1-loop. Here, $G_F^{\mu\nu}$ is the full gauge propagator and $G_F^{(0)~\mu\nu}(q)$ is the same 
at tree-level. We have separated out the contribution due to the covariant $R_{\xi}$ gauge in the last
term, in order to incorporate different forms of tree level propagators through the above equation, 
subjected to different $\mathcal{L}_g$s. On inversion of the above SDE, the full propagator gets contribution
from a class of diagrams made out of a number of vacuum polarization (bubble) terms, making the result
non-perturbative \cite{IZ}.
\paragraph*{}For the gauge Lagrangian in Eq. \ref{1}, with CS term accompanied by the kinetic term
corresponding to the tree-level propagator, 

\be
G_0^{\mu\nu}=-\frac{1}{q^2-\mu^2}\left[\eta^{\mu\nu}-\frac{q^{\mu}q^{\nu}}{q^2}-i\frac{\mu}{q^2}\epsilon^{\mu\nu\rho}q_{\rho}\right]-\xi\frac{q^{\mu}q^{\nu}}{q^4},\label{12a}
\ee
the SDE leads to the 1-loop propagator \cite{Rao,Rao01}, 

\begin{align}
G^{\mu\nu}(q)&\equiv\frac{1}{\left[q^2\{1+\Pi_e(q)\}^2+\left\{\Pi_o(q)+i\mu\right\}^2\right]q^2}\nonumber\\
&\times\left[\left(q^{\mu}q^{\nu}-\eta^{\mu\nu}q^2\right)\{1+\Pi_e(q)\}-\epsilon^{\mu\nu\rho}q_{\rho}\left\{\Pi_o(q)+i\mu\right\}\vphantom{\left(q^{\mu}q^{\nu}-\eta^{\mu\nu}q^2\right)}\right]\nonumber\\
&-\xi\frac{q^{\mu}q^{\nu}}{q^4}.\label{12}
\end{align}
Apart from the usual pole at $q^2=0$, the above propagator has a non-trivial pole, defined by the solution
of the equation,

\be
q^2\{1+\Pi_e(q)\}^2+\left\{\Pi_o(q)+i\mu\right\}^2=0.\label{13}
\ee
The pole of the gauge propagator, if local in the momentum space and gauge-invariant, represents a physical
state of the system \cite{IZ}. Though the latter criteria is satisfied, the prior is in general not true for the 
present case. However, just below the two-fermion threshold, parametrized by $\vert q\vert=2\vert m\vert-\epsilon$,
with $\vert m\vert\gg\epsilon\cong0$ the same is satisfied \cite{H}. However, as this pole appears just below the two-particle
threshold of the integrated-out fermions, this is an effective occurrence of a shallow fermionic {\it bound state}. The
corresponding binding energy (BE) can be identified as,

\be
\epsilon=4\vert m\vert\exp\left\{\frac{4\pi}{e^2}\left(2\vert m\vert-\mu\frac{m}{\vert m\vert}\right)\right\}.\label{14}
\ee
A similar result was obtained by Hagen for pure CS QED, without gauge dynamics, yielding
$\epsilon\approx\exp\left(-4\pi\frac{m}{\vert m\vert}\mu/e^2\right)$ \cite{H}. There, the sign of
fermion mass $\frac{m}{\vert m\vert}$, or the `induced' photon spin \cite{TopM2,TopM22,Boy,Boy1} in 2+1,
has to be positive for $\mu>0$, and negative otherwise, for attaining a sensible (small) value
of $\epsilon$. This is required for validity of the expansion, a fact not stressed upon in Ref. \cite{H}.
It is sensible that the BE disappears for $e^2\rightarrow0$ (weak coupling limit).
\paragraph*{}In the present case, the presence of the kinetic term, representing vacuum fluctuations, opposes
the formation of the bound state, against the CS influence, represented by the exponent in Eq. \ref{14}.
Only beyond a critical value,
\be
\mu\frac{m}{\vert m\vert}\equiv2\vert m\vert:=\mu_c,\label{N01}
\ee
of the CS coefficient, this exponent can 
be negative, provided the photon spin and $\mu$ have the same sign. Then the $\epsilon$-expansion
is meaningful, and thereby corresponds to a shallow bound state. 
\paragraph*{}At the tree-level, the sign of the CS coefficient $\mu$ gives the photon spin.
The quantum effects shift the same as $\mu\rightarrow\mu-i\Pi_o(m,q)$ in the effective gauge 
Lagrangian. Due to the logarithmic singularity of $\Pi_o(m,q)$, at $\vert q\vert\rightarrow2\vert m\vert$,
it dominates $\mu$, yielding the `final' induced photon spin: $m/\vert m\vert$ \cite{Own1}. Physically,
the photon spin at the tree-level is substituted by the combined spin of the fermion-anti-fermion bound state
in the effective theory, the latter ($\frac{m}{2\vert m\vert}$ each) being additive as spin is a $U(1)$
conserved quantity in 2+1 dimensions. To be a bound state, realized in the photon-channel, the bound state
spin must be parallel to the original photon spin ($\mu/\vert\mu\vert$), necessitating the positivity of
$\mu\frac{m}{\vert m\vert}$. In general, the relative sign of $\mu$ and the induced photon spin can be
arbitrary ($\pm1$) near the two-fermion threshold. Only when they have the same sign, the exciton can
exist, provided the condition in Eq. \ref{N01} is satisfied.
\paragraph*{}Fig. \ref{f2} depicts the plot of
Eq. \ref{14} for three different values of $m$ in suitable units. The parametric region corresponding
to physical bound state is observed. 

\subsection{The nature of the bound state}
Since this bound state appears as a pole of the gauge propagator, it is charge-less, and
dissociates into an electron-positron pair at the two particle threshold ($q^2=4m^2$). These assertions
are confirmed by the corresponding 1-loop renormalization coefficients, large-N protected beyond 1-loop,
that can be read-off from Eq. \ref{12}. The Lehmann weight, 

\be
Z_3=\left[1+\Pi_e(q)\right]^{-1}=\left[1+\frac{e^2}{4\pi}\left\{\frac{1}{|q|}\left(\frac{1}{4}+\frac{m^2}{q^2}\right)\log\left(\frac{2\vert m\vert+\vert q\vert}{2\vert m\vert-\vert q\vert}\right)-\frac{|m|}{q^2}\right\}\right]^{-1},\label{N2}
\ee
vanishes
near the two-particle threshold, owing to the logarithmic singularity of Eq. \ref{4}, marking emergence
of bound state \cite{IZ}. Further, the renormalized charge $e^2_r=Z_3e^2$ vanishes in the same 
limit, asserting the state to be charge-less. This still renders the components to be a pair of
particle and anti-particle, as in planar world, both species have the same spin orientation
\cite{TopM1,TopM11,TopM2,TopM22,Boy,Boy1}. Additionally, the renormalized topological mass,

\bea
\mu_r&=&\left[\mu-i\Pi_o(q)\right]Z_3\nonumber\\
&=&\left[\mu-\frac{m}{4\pi}\frac{e^2}{|q|}\log\left(\frac{2\vert m\vert+\vert q\vert}{2\vert m\vert-\vert q\vert}\right)\right]\nonumber\\
&\times&\left[1+\frac{e^2}{4\pi}\left\{\frac{1}{|q|}\left(\frac{1}{4}+\frac{m^2}{q^2}\right)\log\left(\frac{2\vert m\vert+\vert q\vert}{2\vert m\vert-\vert q\vert}\right)-\frac{|m|}{q^2}\right\}\right]^{-1},\label{N3}
\eea
leads to $\mu_r^2\ge4m^2$ near the two-particle threshold, reflecting the threshold condition for
the formation of the bound state.   

\paragraph*{}The topological origin of this bound state can be analogous to the finite boundary
of the K-space of a superconductor, leading the expression of the corresponding gap. In the latter case,
the order parameter $\Delta$ is introduced as an auxiliary interaction field, marking an additional
energy scale in the system, that physically represents the BE of the Cooper-pair
\cite{FW}. However, this requires an underlying non-relativistic physics, represented by
the Ginzburg-Landau (GL) theory \cite{FW}, unlike the present case. Further, the critical
temperature of the BCS theory satisfies $T_c\propto\vert\Delta\vert$, which is not the case 
here, as it will be shown that a melting temperature of the topological bound state cannot be evaluated
exactly.  
\paragraph*{} From Eq. \ref{12a}, the tree-level gauge theory has mass equal to the CS coefficient
itself \cite{TopM1,TopM11,TopM2,TopM22,Red,Red1}. The quantum effects shift that pole to accommodate a bound
state, with a lower-bound $\mu_c=2\vert m\vert$ on the topological mass magnitude (Eq. \ref{14}). As the
form factors are of $\mathcal{O}(e^2)$, expansion of Eq. \ref{13} yields,

\be
\mu^2=q^2+\frac{2}{\vert m\vert}\left[\vert m\vert\Pi_e+i\mu m\Pi_o\right],~~~\Pi_o\in\Im,~~\Pi_e\in\Re\nonumber
\ee
As the term in the square bracket is a positive definite [Fig. \ref{f1}], $\mu^2>q^2$ and near the
two-particle threshold, $\mu^2>4m^2$. The bound-state vanishes also for $\mu^2\gg4m^2$,
as $\epsilon\rightarrow0$.
  
\subsection{Duality}
The physical reason as to why the CS term allows for the bound state pole in the photon
channel is of critical interest. More interestingly, destabilization effect due to vacuum fluctuations
does not entirely eliminate the same, but only introduces a finite threshold for formation. This can be
understood from the {\it duality} of the present theory with the massive CS vector field \cite{Equi}, 

\be
\mathcal{L}_g=\mu^2a^{\mu}a_{\mu}+\frac{\mu}{2}\epsilon^{\mu\nu\rho}a_{\mu}\partial_{\nu}a_{\rho},\nonumber
\ee
with $\mu$ being the mass. They both yield essentially the same equation of motion, and are different
constrained forms of a more extensive Lagrangian \cite{Equi}. The corresponding tree-level propagators
are related as,
\bea
iq^{\beta}\epsilon_{\beta\gamma\nu}\left(\tilde{G}_0^{-1}\right)^{\mu\nu}&=&\mu\left[\left(G_0^{-1}\right)^{\mu}_{~\gamma}+\frac{1}{\xi}q^\mu q_{\gamma}\right]~~~\text{and}\nonumber\\
-i\epsilon_{\sigma\beta\mu}q^\beta\left(G_0^{-1}\right)^{\mu\nu}&=&\frac{q^2}{\mu}\left(\tilde{G}_0^{-1}\right)^{~\nu}_\sigma-\mu q^\nu q_\sigma,\label{14a}
\eea
where $\tilde{G}_0$ corresponds to the non-dynamic CS QED, expressed as,

\be
\tilde{G}_0^{\mu\nu}=-\frac{1}{q^2-\mu^2}\left[\eta^{\mu\nu}-\frac{1}{\mu^2}q^{\mu}q^{\nu}-i\frac{1}{\mu}\epsilon^{\mu\nu\rho}q_{\rho}\right].\label{14a1}
\ee
The corresponding Lagrangians are connected through a gauge transformation, singular for
$\mu=0$ \cite{Mukhi}. This explains the singularity of $\tilde{G}_0^{\mu\nu}$ and the extra
longitudinal degree of freedoms appearing in Eqs. \ref{14a}. This tree level duality is
robust to quantum corrections \cite{Equi} in presence of interaction generating vacuum
polarization, as,

\bea
0&=&\left(\mu^2-q^2\Pi_e\right)^2+\left(\Pi_o-i\mu\right)^2q^2,~~~\text{yielding},\nonumber\\
\epsilon&=&4\vert m\vert\exp\left[\frac{2\pi\mu}{e^2\vert m\vert}(2m-\mu)\right].\label{14b2}
\eea
Though the pole equation is different, the condition for attaining a bound state is exactly the
same as dynamic CS QED ($\mu^2>4m^2$). This can also be obtained through the transformations
among the respective 1-loop corrected propagators, as in Eq. \ref{14a}. This further confirms
the physicality of the bound-state pole, as the `duality' is essentially a contraction of the
inverse propagator. Additionally, for the massive case, inclusion of quantum corrections still
maintains the need of gauge-fixing {\it after} the gauge transformation.  
\paragraph*{} Thus, the topologically massive gauge theory is equivalent to a
genuinely massive pure topological theory in terms of the bound state. Exploiting this,
as is clearer in the latter case, whenever the photon mass is above the two-particle threshold of
the fermions, it transforms into electron-positron pair. When it is slightly below that
(owing to quantum corrections), there is no pair production and the difference is the
BE. This explains the lower-bound $\mu_c$ of the coefficient of the topological
term magnitude in the dynamic CS-QED to be same as the physical mass of the dual massive theory at
two-particle threshold.

\section{Effect of Thermal Fluctuations}
We adopt the imaginary time formalism \cite{DasFT} to analyze the behavior of this bound state 
at finite temperature. The evaluation for vacuum polarization in this formalism, for $QED_3$
(planar QED Wick-rotated into the Euclidean space) was done for {\it massless} fermions in Ref.
\cite{Dorey}. We extend it to the case of $m\neq0$, in a form more adaptive to our nomenclature,
and show that the already known specific results, both at finite and zero-temperature are different
limiting cases of the present case.
\paragraph*{} To this end, we adopt the following definition of vacuum polarization \cite{Dorey}, 

\bea
\Pi^{\mu\nu}(q)&=&ie^2Tr_D\int_p\gamma^{\mu}S_F(p)\gamma^{\nu}S_F(p-q),\nonumber\\
S_F(p)&:=&\left[\gamma^{\mu}p_{\mu}-m\right],~~~p_{\mu}:=i\partial_{\mu},\label{28}
\eea
where, in contrast with Eq. \ref{3}, we have left out the Schwinger regularization term for brevity.
Upon the Wick rotation to the imaginary time, the fermion variables are,

\begin{eqnarray}
p_E&=&(p_3,\vec{p}),~~~p_3=(2n+1)\frac{\pi}{\beta},\nonumber\\
p^2_3&=&-p^2_0,~~~p^2_E=p^2_3+\vec{p}^2=-p^2,\nonumber\\
n&=&0,\pm1,\pm2\ldots,~~~\beta=1/T,
\end{eqnarray}
and the boson variables are,

\begin{eqnarray}
q_E&=&(q_3,\vec{q}),~~~q_3=2r\frac{\pi}{\beta},\nonumber\\
q^2_3&=&-q^2_0,~~~q^2_E=q^2_3+\vec{q}^2=-q^2\nonumber\\
r&=&0,\pm1,\pm2\ldots,
\end{eqnarray}
The manifest co-variant projections of vacuum polarization tensor, in Euclidean space, are introduced
as \cite{DasFT},

\begin{eqnarray}
\Pi^{\mu\nu}(q_E,\beta)&=&\Pi_A(q_E,\beta)A^{\mu\nu}(q_E)+\Pi_B(q_E,\beta)B^{\mu\nu}(q_E),\nonumber\\
A^{\mu\nu}(q_E)&:=&\left(\delta^{\mu3}-\frac{q^{\mu}q^3}{q^2_E}\right)\frac{q^2_E}{\vec{q}^2}\left(\delta^{3\nu}-\frac{q^3q^{\nu}}{q^2_E}\right),\nonumber\\
B^{\mu\nu}(q_E)&:=&\delta^{\mu i}\left(\delta^{ij}-\frac{q^iq^j}{\vec{q}^2}\right)\delta^{j\nu},\nonumber\\
A^{\mu\nu}+B^{\mu\nu}&\equiv&\delta^{\mu\nu}-\frac{q^{\mu}q^{\nu}}{q^2_E}\equiv\frac{1}{q^2}Q^{\mu\nu},~~~\delta^{\mu\nu}=-\eta^{\mu\nu}\nonumber
\end{eqnarray}
The last line of above equations leads to the physical constraint that, at $T=0$,

\be
\Pi_A(q_E,T=0)=\Pi_B(q_E,T=0)\equiv\Pi_e(q)\label{29}
\ee
From the definitions, it is obtained that,

\bea
\Pi_A(q_E,\beta)&=&\frac{q^2_E}{\vec{q}^2}\Pi^{00}(q_E,\beta)~~\text{and}\nonumber\\
\Pi_B(q_E,\beta)&=&-\Pi^{ii}(q_E,\beta)-\frac{q^2_3}{\vec{q}^2}\Pi^{00}(q_E,\beta),\label{30}
\eea
where repeated indices mean summation, unless mentioned otherwise.
\paragraph*{} Therefore it suffices to evaluate the temporal and spatial components of
vacuum polarization tensor. First we will obtain those for the parity-even part, which is,

\begin{align}
\Pi^{\mu\nu}_e(q)&\equiv i2e^2\int_0^1dx\int_p\frac{1}{\left(p^2-a^2\right)^2}\left[2p^{\mu}p^{\nu}+2x(1-x)\left(q^\mu q^\nu-\eta^{\mu\nu}q^2\right)\right.\nonumber\\
&\qquad~~~~~~~~~-\left.\eta^{\mu\nu}\left(p^2-a^2\right)-(2x-1)\left(\eta^{\mu\nu}p.q-p^{\mu}q^{\nu}-p^{\nu}q^{\mu}\right)\right],\nonumber
\end{align}
where we have utilized the Feynman integration trick with the shift $p\rightarrow p+xq$, and upon 
continuing to Euclidean space by the rules:

\bea
p_0&=&ip_3,~~~p^2=-p^2_E,~~~q^2=-q^2_E,~~~p.q=-p_E.q_E,\nonumber\\
Q^{\mu\nu}&=&Q^{\mu\nu}_E=\delta^{\mu\nu}q^2_E-q^\mu q^\nu,~~~a^2=a^2_E=m^2+x(1-x)q^2_E,\nonumber\\
\int\frac{dp_3}{2\pi}&\rightarrow&\frac{1}{\beta}\sum_{n=-\infty}^{\infty},~~~p_3=\frac{2\pi}{\beta}\left(n+X\right),~~~X:=\frac{1}{2}+xr.\label{31}
\eea
With these definitions, one can express the temporal and spatial even form factors as,

\begin{align}
\Pi^{00}(q_E,\beta)&=-2\frac{e^2}{\beta}\int_0^1dx\int\frac{d^2p}{(2\pi)^2}\left[S_1-2p^2_3S_2-2x(1-x)\vec{q}^2S_2-(2x-1)q_3S^*\right],~~~\text{\&}\nonumber\\
\Pi^{ii}(q_E,\beta)&=-2\frac{e^2}{\beta}\int_0^1dx\int\frac{d^2p}{(2\pi)^2}\left[2x(1-x)\left(q^2_E+q^2_3\right)S_2-2a_E^2S_2-2p^2_3S_2+(2x-1)q_3S^*\right];\nonumber\\
S_i&=\sum_{n=-\infty}^{\infty}\frac{1}{\left[p^2_E+a^2_E\right]^i},~~i=1,2~~~\text{\&}~~~S^*=\sum_{n=-\infty}^{\infty}\frac{p_3}{\left[p^2_E+a^2_E\right]^2}.\nonumber
\end{align}
For the massive fermion, the frequency sums followed by 2-momentum integrals yields,

\begin{align}
\int\frac{d^2p}{(2\pi)^2}S_1&=-\frac{1}{4\pi}\log\left\vert\frac{\sin\pi\left(X+\frac{i\beta m}{2\pi}\right)}{\sin\pi\left(X+\frac{i\beta a_E}{2\pi}\right)}\right\vert,~~~\int\frac{d^2p}{(2\pi)^2}S_2=\frac{\beta}{16\pi}\frac{1}{a_E}\Im\cot\pi\left(X+\frac{i\beta a_E}{2\pi}\right),\nonumber\\
\int\frac{d^2p}{(2\pi)^2}S^*&=\frac{\beta}{16\pi}\left[\sin(2\pi X)-\Re\cot\pi\left(X+\frac{i\beta a_E}{2\pi}\right)\right],\nonumber\\
\int\frac{d^2p}{(2\pi)^2}p_3^2S^*&=\frac{1}{4}\Im\left[\frac{\sin\pi\left(X+\frac{i\beta m}{2\pi}\right)\cos\pi\left(X-\frac{i\beta m}{2\pi}\right)}{\left\vert\sin\pi\left(X+\frac{i\beta m}{2\pi}\right)\right\vert^2}+\cot\pi\left(X+\frac{i\beta a_E}{2\pi}\right)\right].\label{31}
\end{align}

Finally, from Eq. \ref{29}, the finite temperature `orthogonal' form factors \cite{DasFT}
are obtained as,

\begin{align}
\Pi_A&=\frac{e^2q^2_E}{\beta\vec{q}^2}\int_x\left[\frac{1}{2\pi}\log\left\vert\frac{\sin\pi\left(X^+_m\right)}{\sin\pi\left(X^+_a\right)}\right\vert-\Im\left\{\cot\pi\left(X^-_m\right)+\cot\pi\left(X^+_a\right)\right\}\right]\nonumber\\
&+\frac{e^2}{8\pi}\int_x\left[2x(1-x)\frac{q^2_E}{a_E}\Im\cot\pi\left(X^+_a\right)+(2x-1)q_3\frac{q^2_E}{\vec{q}^2}\left\{\sin(2\pi X)-\Re\cot\pi\left(X^+_a\right)\right\}\right]~~~\text{\&}\nonumber\\
\Pi_B&=-\frac{e^2}{\beta}\int_x\left[\frac{1}{2\pi}\log\left\vert\frac{\sin\pi\left(X^+_m\right)}{\sin\pi\left(X^+_a\right)}\right\vert-\frac{q^2_E}{\vec{q}^2}\Im\left\{\cot\pi\left(X^-_m\right)+\cot\pi\left(X^+_a\right)\right\}\right]\nonumber\\
&+\frac{e^2}{8\pi}\int_x\left[-2\frac{m^2}{a_E}\Im\cot\pi\left(X^+_a\right)+(2x-1)q_3\left(2-\frac{q^2_3}{\vec{q}^2}\right)\left\{\sin(2\pi X)-\Re\cot\pi\left(X^+_a\right)\right\}\right];\nonumber\\
X^\pm_m&=X\pm\frac{i\beta m}{2\pi}~~~\text{and}~~~X^\pm_a=X\pm\frac{i\beta a_E}{2\pi}.\label{32}
\end{align}
Here,
the Schwinger term regularization has been adopted as appropriate. The $x$-integrals cannot be evaluated
exactly, as is known. However, at zero temperature, the same is possible, and one obtains
$\Pi_A=\Pi_B$ as required; modulo an additional term in $\Pi_B$, known to arise for
fermion-antifermion pair of the loop having the same mass \cite{DasFT}. The rest is same
as the result given in Eq. \ref{4}, once the expressions are continued back to the
Minkowski space through $q_3\rightarrow-iq_0$. 
Additionally, in Eqs. \ref{32}, terms with overall multiplicative factor $T=1/\beta$ have been separated
from the rest for convenience. This is because, confirmed by a analysis in real time formalism, the
$x$-integrals turn out to be temperature-independent in the high-temperature limit, as will be seen
in the next subsection. Then, the dominant temperature dependent and independent parts become well-separated.
\paragraph*{}As for the Chern-Simons coefficient, following a similar treatment leads to,

\be
\Pi_o(q_E,\beta)\equiv i\frac{me^2}{4\pi}\int_0^1\frac{dx}{a_E}\Im\cot\pi\left(X+is\frac{\beta a_E}{2\pi}\right)\label{33}
\ee 
In the zero-temperature limit, the above expression goes to that of Eqs. \ref{4} as required.
Further, in the high temperature limit ($\beta\rightarrow0$), the dominant $\mathcal{O}(\beta)$
term is independent of $q$, and is the same as that for $q=0$ \cite{BDP}. These facts ensure
the consistency of our derivation. 

\subsection{Results at high temperatures}
\begin{figure}
\centering 
\includegraphics[width=5 in]{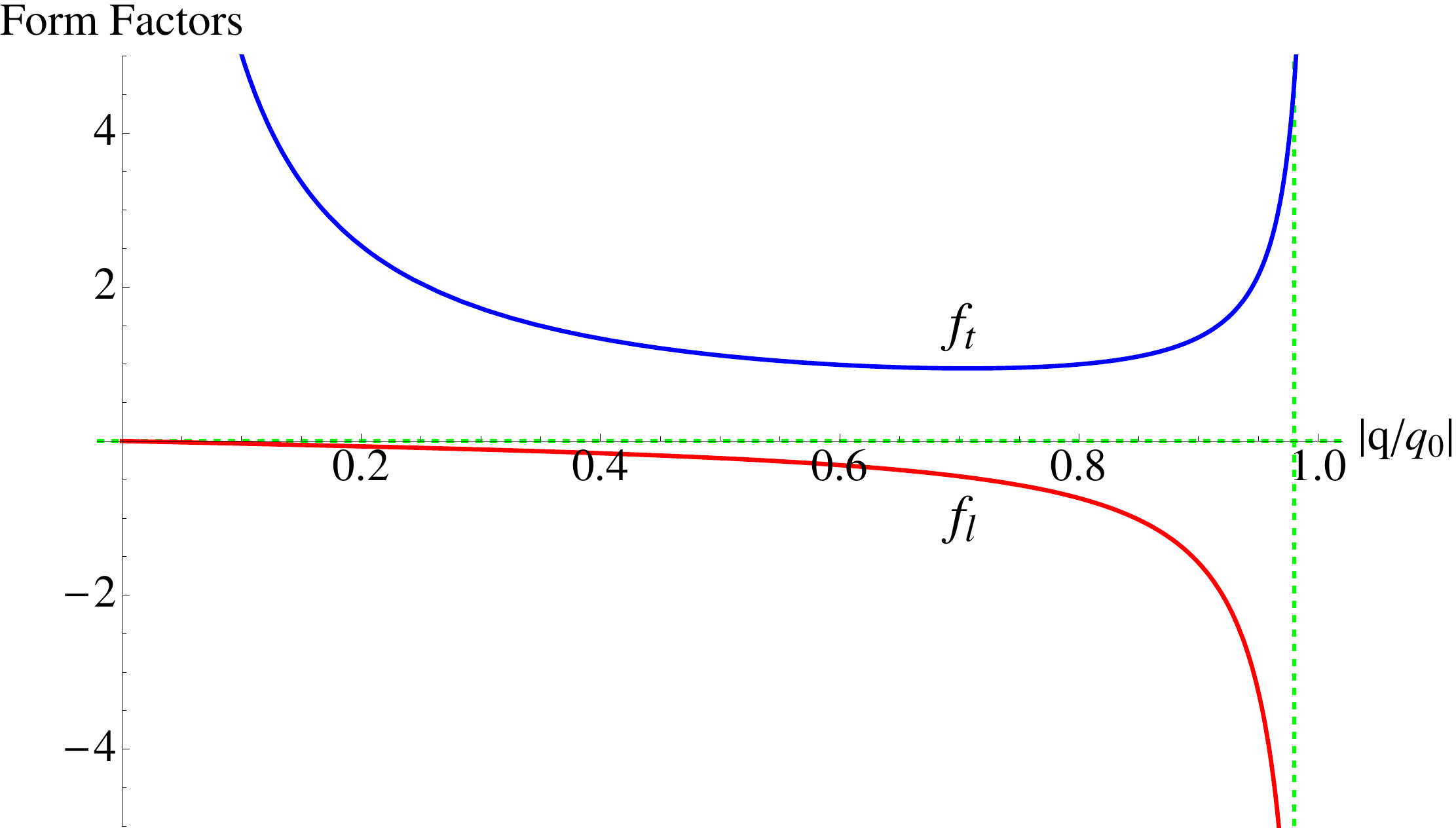}
\caption{Plots of finite-temperature parts of coefficients of vacuum polarization
tensor Vs. $\vert q/q_0\vert$. Here, $f_{l,t(q)}=2\pi\Pi_{l,t}(q)/\log(2)$. $f_t(q)$ is positive and hence
is acceptable, within the physical region $0\le\vert q/q_0\vert\le1$, whereas $f_l(q)$ is not.}\label{f3}
\end{figure}

Though the $x$-integrals in Eqs. \ref{32} and \ref{33} cannot be evaluated exactly, approximate 
expressions, in the high temperature limit $T\gg q,m$ \cite{TFT}, can be obtained as,

\bea
\Pi_A(q,T)&\approx&\frac{T}{2\pi}e^2\log(2)\frac{\vert \frac{q}{q_0}\vert}{1-\vert\frac{q}{q_0}\vert^2}\left[1-\frac{2}{3}\frac{\vert \frac{q}{q_0}\vert^2}{1-\vert\frac{q}{q_0}\vert^2}\right]-q^2\Pi_e(q),\nonumber\\
\Pi_B(q,T)&\approx&-\frac{T}{3\pi}e^2\log(2)\frac{\vert \frac{q}{q_0}\vert^3}{1-\vert\frac{q}{q_0}\vert^2}-q^2\Pi_e(q),~~~\Pi_o(q,T)\approx -i\frac{\beta e^2}{8\pi}m.\label{23}
\eea
The even part of the vacuum polarization at finite temperature can be expressed as \cite{Weldon},

\bea
\Pi^{\mu\nu}_e(q,T)&=&\Pi_T(q,\omega)P^{\mu\nu}+\Pi_L(q,\omega)R^{\mu\nu},\label{16}\\
P^{\mu\nu}&=&\eta^{\mu\nu}-u^{\mu}u^{\nu}+\frac{\tilde q^{\mu}\tilde q^{\nu}}{Q^2},~~~R_{\mu\nu}=-\frac{1}{q^2Q^2}(Q^2u_{\mu}+\omega\tilde q_{\mu})(Q^2u_{\nu}+\omega\tilde q_{\nu}),\nonumber\\
\omega&=&q.u,~~~Q^2=\omega^2-q^2,~~~\tilde q^{\mu}=q^{\mu}-\omega u^{\mu}.\nonumber
\eea
with $\Pi_T=\Pi_A+q^2\Pi_e$ and $\Pi_L=\Pi_B+q^2\Pi_e$. here, $u_\mu=(1,0,0)$ is the position 
vector defining the rest frame of the thermal bath \cite{DasFT}. A straight-forward, but tedious 
calculation leads to the 1-loop corrected full thermal propagator as,

\bea
G^{\mu\nu}(q,T)&\equiv& a(q,T)\eta^{\mu\nu}+b(q,T)q^{\mu}q^{\nu}+c(q,T)u^{\mu}u^{\nu}\nonumber\\
&+&d(q,T)(q^{\mu}u^{\nu}+u^{\mu}q^{\nu})+e(q,T)\epsilon^{\mu\nu\rho}q_{\rho};\nonumber\\
a(q,T)&=&\frac{\Pi_T-q^2\left(\Pi_e+1\right)}{\left[\Pi_T-q^2\left(\Pi_e+1\right)\right]^2+q^2\left(\Pi_o+i\mu\right)^2},\nonumber\\
b(q,T)&=&\frac{\left[\Pi_L-q^2\left(\Pi_e+1\right)\right]+\frac{q_0^2}{\vec q^2}(\Pi_T-\Pi_L)}{q^2\left[q^2\left(\Pi_e+1\right)-\Pi_L\right]}a(q,T)+\frac{\xi}{q^4},\nonumber\\
c(q,T)&=&-\frac{q^2}{\vec q^2}\frac{(\Pi_T-\Pi_L)}{\left[\Pi_L-q^2\left(\Pi_e+1\right)\right]}a(q,T),\nonumber\\
d(q,T)&=&-\frac{q_0}{q^2}c(q,T),~~~e(q,T)=-\frac{\left(\Pi_o+i\mu\right)}{\Pi_T-q^2\left(\Pi_e+1\right)}a(q,T),\label{21}
\eea
The above equations reveal {\it two} non-trivial poles of the full propagator, corresponding to the 
identities,

\bea
\Pi_L-q^2\left(\Pi_e+1\right)&=&0~~~\text{and}\nonumber\\
\left[\Pi_T-q^2\left(\Pi_e+1\right)\right]^2+q^2\left(\Pi_o+i\mu\right)^2&=&0.\label{22}
\eea
Near the two-fermion threshold, the above identities corresponds to the respective bound state binding
energies as,

\bea
\epsilon&\approx&4\vert m\vert\exp{\left\{-{\rm T}\frac{4\pi}{\vert m\vert}\Pi_l+16\pi\frac{\vert m\vert}{e^2}\right\}}~~~\text{\&}\nonumber\\
\epsilon&\approx&4\vert m\vert\exp{\left\{\frac{4\pi}{e^2}\left(2\vert m\vert-\mu\frac{m}{\vert m\vert}\right)-8\pi{\rm T}\Pi_t\left(2\vert m\vert+\mu\frac{m}{\vert m\vert}\right)^{-1}\right\}},\label{27}
\eea
where $\Pi_{l,t}(q)=\Pi_{L,T}(q,{\rm T})/{\rm T}e^2$. Here, we have retained the ${\cal O}(e^0)$ term, along-with
the dominant ${\cal O}(e^{-2})$ contribution, as the prior carries the temperature-dependence. In contrast,
only the ${\cal O}(e^{-2})$ terms were retained in the exponent for the zero-temperature case (Eq. \ref{14}).
However, the temperature-dependent part can substantially contribute, as the form factor $\Pi_t$ can be large,
as can be seen in Fig. \ref{f3}. In the above, the first expression is unphysical, as in the
on-shell domain of $\vert q\vert<\vert q_0\vert$, $\Pi_l$ is negative, marking increase in binding
energy with temperature. However, the second one is physically sensible, as the scaled form factor
$\Pi_t$ is positive on-shell, representing melting of the bound state (Fig. \ref{f3}). This, however, 
requires the factor $\mu m/\vert m\vert$ to be positive, required for a physically sensible bound-state,
as had been seen for the zero-temperature case. Further, this expression
reverts back to the result in Eq. \ref{14} for $T=0$, as physically expected. This makes sense, although
the expression for $\Pi_t$ has been obtained in the high-temperature limit, where it is temperature-independent.
Though its exact form is expected to show temperature-dependence, the temperature dependent parts of the
functions in the integrands of Eqs. \ref{32} (the logarithms and cotangents) vanish identically for $T=0$,
justifying the recovery of Eq. \ref{14}. Further, the finite temperature contribution is expected to be
finite for any finite value of $T$, as finite temperature effects do not introduce additional singularities
to the overall form factors \cite{DasFT}. Therefore, ${\rm T}\Pi_t$ should vanish for $T=0$.

\paragraph*{} Therefore, the finite temperature essentially does an exponential scaling of the 
electron-positron BE, under the high-$T$ approximation. Presently, the `critical'
temperature at which the bound state melts is not that straight-forward to obtain from this
approximate approach. This exponential decay represents the destabilizing thermal fluctuations,
as mentioned before, that lowers the BE than what it was at $T=0$. This shift may
be observable in a physical system.
\paragraph*{} It has been shown that \cite{Grig} 2+1 QED (without a tree-level CS term) shows
confinement-deconfinement transition of Berezinskii-Kosterlits-Thouless (BKT) \cite{BKT,BKT1} type.
However, such a theory cannot account for a parity violating mass term and thus it emerges
from the consideration of $4\times4$ Dirac matrices, unlike the $2\times2$ Pauli matrices
as gamma matrices in the Dirac equation. This allows cancellation between the CS terms at the
loop levels corresponding to particle and anti-particle sectors. Thus, it is imperative to
note that in theories with topological terms and loop corrections
are analyzed, possible phase transitions are not of BKT type. This also follows from the
fact that BKT is an infinite order phase transition, unlike BCS phase transition, which is 
of second order and has analytic correspondence with the present theory.

\section{Discussions and Remarks}
The presence of both quantum and thermal fluctuations for gauge field destabilizes
the fermion-antifermion bound state, with respect to the one for pure CS QED. The
lower limit for the CS coefficient can signify a quantum `phase-transition'. It
will be interesting to study the non-Abelian analogue of this model, as it
also supports anomalies. The finite temperature treatment was demonstrated for massive fermions, which
is consistent with the physical understanding of melting of the bound state. A first-quantized treatment
of the model will be discussed in a following body of work, supporting complex angular momentum,
resulting in Efimov-state-like resonances.
\paragraph*{}The fact that planar QED has manifested in effective low-energy theories of materials like graphene \cite{Own1},
with $U(1)$ gauge fields \cite{Ando01} and controlled mass gaps \cite{Hunt}. The mono-layer graphene is 
described by relativistic fermions \cite{Gra2}, whereas the bi-layer one is governed by non-relativistic
dynamics \cite{Neto}. Formation of bound states in these systems and their controlled manipulation \cite{Own1} is
of deep interest in the physical context. The manifest relativistic physics arise through emergent
Dispersion in graphene \cite{Gra2}, wherein the fermion mass and coupling strength get scaled,
respectively, as $m\rightarrow mv_F^2$ and $e\rightarrow ev_F\ll c$ \cite{Own1} by the Fermi velocity
$v_F$, replacing the speed of light in vacuum ($c$) as a Lorentz invariant. The latter scaling makes the 
BE shallower, thereby increasing the sensitivity of the bound state to the parametric threshold. Further,
the melting temperature of the bound state should not be lower than the range where the structure of these
materials is intact, for the physical observation of the prior. 

\acknowledgments
The authors appreciate Dr. Vivek M. Vyas for valuable inputs.
KA is thankful to Prof. Ashok Das for many useful suggestions.

\end{document}